\documentclass[conference]{IEEEtran}
\IEEEoverridecommandlockouts

\usepackage{url}
\usepackage{xcolor}

\usepackage{hyperref}

\usepackage{import}
\usepackage{cite}
\usepackage{amsmath,amssymb,amsfonts}
\usepackage{latexsym}
\usepackage{algorithmic}
\usepackage{graphicx}
\usepackage{textcomp}
\usepackage{xcolor}
\usepackage{tikz}
\usepackage{tabularx}
\usepackage{multirow}
\usepackage{{booktabs}}
\usepackage{longtable}
\usepackage[acronym]{glossaries}
\definecolor{newcolor}{rgb}{.8,.349,.1}
\def\BibTeX{{\rm B\kern-.05em{\sc i\kern-.025em b}\kern-.08em
		T\kern-.1667em\lower.7ex\hbox{E}\kern-.125emX}}

\newacronym{ai}{AI}{Artificial Intelligence}
\newacronym{ml}{ML}{Machine Learning}
\newacronym{dl}{DL}{Deep Learning}
\newacronym{cadx}{CADx}{Computer-Aided Diagnosis}
\newacronym{ct}{CT}{Computed Tomography}
\newacronym{mri}{MRI}{Magnetic Resonance Imaging}
\newacronym{pet}{PET}{Positron Emission Tomography}
\newacronym{cade}{CADe}{Computer-aided detection}
\newacronym{fda}{FDA}{US Food and Drug Administration}
\newacronym{gui}{GUI}{Graphical User Interface}
\newacronym{prisma}{PRISMA}{Preferred Reporting Items for Systematic Reviews and Meta-Analyses}
\newacronym{oct}{OCT}{Optical Coherence Tomography}
\newacronym{cnn}{CNN}{Convolutional Neural Network}
\newacronym{cdr}{CDR}{Cup-to-Disc Ratio}
\newacronym{svm}{SVM}{Support Vector Machine}
\newacronym{knn}{KNN}{k-Nearest Neighbor}
\newacronym{rf}{RF}{Random Forest}
\newacronym{dt}{DT}{Decision Tree}
\newacronym{rof}{RF}{Rotation Forest}
\newacronym{nb}{NB}{Naïve Bayes}
\newacronym{ann}{ANN}{Artificial Neural Network}
\newacronym{mlp}{MLP}{Multilayer Perceptron}
\newacronym{lr}{LR}{Logistic Regression}
\newacronym{pca}{PCA}{Prinicipal Component Analysis}
\newacronym{lda}{LDA}{Linear Discriminant Analysis}
\newacronym{qda}{QDA}{Quadratic Discriminant Analysis}
\newacronym{mrmr}{MRMR}{Minimum Redundancy Maximum Relevance}
\newacronym{od}{OD}{Optic Disc}
\newacronym{oc}{OC}{Optic Cup}
\newacronym{slo}{SLO}{Scanning Laser Ophthalmoscopy}
\newacronym{lsfg}{LSFG}{Laser Speckle Flowgraphy}
\newacronym{fa}{FA}{Fluorescein Angiography}
\newacronym{vf}{VF}{Visual Field}
\newacronym{vit}{ViT}{Vision Transformers}
\newacronym{lstm}{LSTM}{Long Short-Term Memory}
\newacronym{rnfl}{RNFL}{Retinal Nerve Fiber Layer}
\newacronym{onh}{ONH}{Optic Nerve Head}
\newacronym{lbp}{LBP}{Local Binary Pattern}
\newacronym{lcp}{LCP}{Local Configuration Pattern}
\newacronym{glrm}{GLRM}{Grey Level Run Length Matrix}
\newacronym{glcm}{GLCM}{Grey Level Co-occurrence Matrix}
\newacronym{hog}{HOG}{Histogram of Oriented Gradients}
\newacronym{hos}{HOS}{Higher Order Spectra}
\newacronym{hoc}{HOC}{Higher Order Cumulant}
\newacronym{lsda}{LSDA}{Locality Sensitive Discriminant Analysis}
\newacronym{sffs}{SFFS}{Sequential Floating Forward Search}
\newacronym{gan}{GAN}{Generative Adversarial Network}
\newacronym{rnn}{RNN}{Recurrent Neural Network}
\newacronym{auroc}{AUROC}{Area Under Receiver Operating Characteristic Curve}
\newacronym{auc}{AUC}{Area Under Curve}
\newacronym{sen}{SEN}{Sensitivity}
\newacronym{spe}{SPE}{Specificity}
\newacronym{acc}{ACC}{Accuracy}
\newacronym{f1eq}{F_{1}}{F1-Score}
\newacronym{f1}{$\acrshort{f1eq}$}{F1-Score}
\newacronym{tpr}{TPR}{True Positive Rate}
\newacronym{fpr}{FPR}{False Positive Rate}
\newacronym{p}{P}{Positive}
\newacronym{n}{N}{Negative}
\newacronym{pp}{PP}{Positive}
\newacronym{pn}{PN}{Negative}
\newacronym{tp}{TP}{True Positive}
\newacronym{tn}{TN}{True Negative}
\newacronym{fp}{FP}{False Positive}
\newacronym{fn}{FN}{False Negative}
\newacronym{capsnet}{CapsNet}{Capsule Network}
\newacronym{roi}{ROI}{Region of Interest}
\newacronym{cv}{CV}{Computer Vision}
\newacronym{isnt}{ISNT}{Inferior, Superior, Nasal, and Temporal}
\newacronym{ddls}{DDLS}{Disc Damage Likelihood Scale}
\newacronym{poag}{POAG}{Primary Open-Angle Glaucoma}
\newacronym{pacg}{PACG}{Primary Angle-Closure Glaucoma}
\newacronym{2d_fbse_ewt}{2D-FBSE-EWT}{Two-dimensional Fourier-Bessel Series Expansion-empirical Wavelet Transform}
\newacronym{gsm_dcn}{GSM-DCM}{Glaucoma Syndrome Mechanism-based Dual-Channel Network}

\begin{document}
	
\title{AI-Driven Approaches for Glaucoma Detection - A Comprehensive Review}
	

\author{
    \IEEEauthorblockN{Yuki Hagiwara\IEEEauthorrefmark{1}, 
    Octavia-Andreea Ciora\IEEEauthorrefmark{1}, 
    Maureen Monnet\IEEEauthorrefmark{1}, 
    Gino Lancho\IEEEauthorrefmark{1}, 
    Jeanette Miriam Lorenz\IEEEauthorrefmark{1}\IEEEauthorrefmark{2}}
    \IEEEauthorblockA{\IEEEauthorrefmark{1}Fraunhofer Institute for Cognitive Systems IKS, Munich, Germany}
    \IEEEauthorblockA{\IEEEauthorrefmark{2}Ludwig Maximilian University, Faculty of Physics, Munich, Germany}
}

\maketitle
	
\begin{abstract}
The diagnosis of glaucoma plays a critical role in the management and treatment of this vision-threatening disease. Glaucoma is a group of eye  diseases that cause blindness by damaging the optic nerve at the back of the eye. Often called "silent thief of sight", it exhibits no symptoms during the early stages. Therefore, early detection is crucial to prevent vision loss. With the rise of Artificial Intelligence (AI), particularly Deep Learning (DL) techniques, Computer-Aided Diagnosis (CADx) systems have emerged as promising tools to assist clinicians in accurately diagnosing glaucoma early. This paper aims to provide a comprehensive overview of AI techniques utilized in CADx systems for glaucoma diagnosis. Through a detailed analysis of current literature, we identify key gaps and challenges in these systems, emphasizing the need for improved safety, reliability, interpretability, and explainability. 
By identifying research gaps, we aim to advance the field of CADx systems especially for the early diagnosis of glaucoma, in order to prevent any potential loss of vision. 
\end{abstract}


\section{Introduction}
Glaucoma is a group of eye diseases that can lead to blindness due to the damage to the optic nerve at the back of the eye. It is commonly associated with increased intraocular pressure~\cite{VARMA_assessment_glaucoma}. This elevated pressure pushes against the optic nerve, damaging its fibers, which in turn leads to the deterioration of the \gls{rnfl} and results in an enlarged \gls{cdr} or \gls{onh}. Hence, a greater \gls{cdr} is often associated with glaucoma. The \gls{cdr} is typically determined via the ratio of the vertical diameter of the \gls{oc} to the vertical diameter of the \gls{od} \cite{ahmad_glaucoma_2018}.   
It is the leading cause of irreversible blindness. According to the 2024 Glaucoma Report, it is estimated that currently, 80 million people globally have glaucoma \cite{geller_glaucoma_stats_2024}. It is estimated that 111.8 million people will be affected by glaucoma in 2040 \cite{tham_global_2014}. 
It is often referred to as the "silent thief of sight" as it exhibits no symptoms during the early stages. Therefore, it is crucial to detect glaucoma early \cite{wiggs_glaucoma_2014}. However, it is challenging to detect glaucoma during its early stages. Many factors contribute to this difficulty, including the subtle nature of early symptoms and the reliance on nuanced interpretations of retinal images. Doctors, especially those who are new or less experienced, may find it challenging to make accurate diagnoses \cite{chicago_diagnosing_glaucoma_2018}. Moreover, diagnoses can be subjective which could lead to inconsistencies and delays in identifying the disease. 

\begin{figure*}[!h]
\centering
\includegraphics[trim={0 3cm 0 2cm},clip, width=\textwidth]{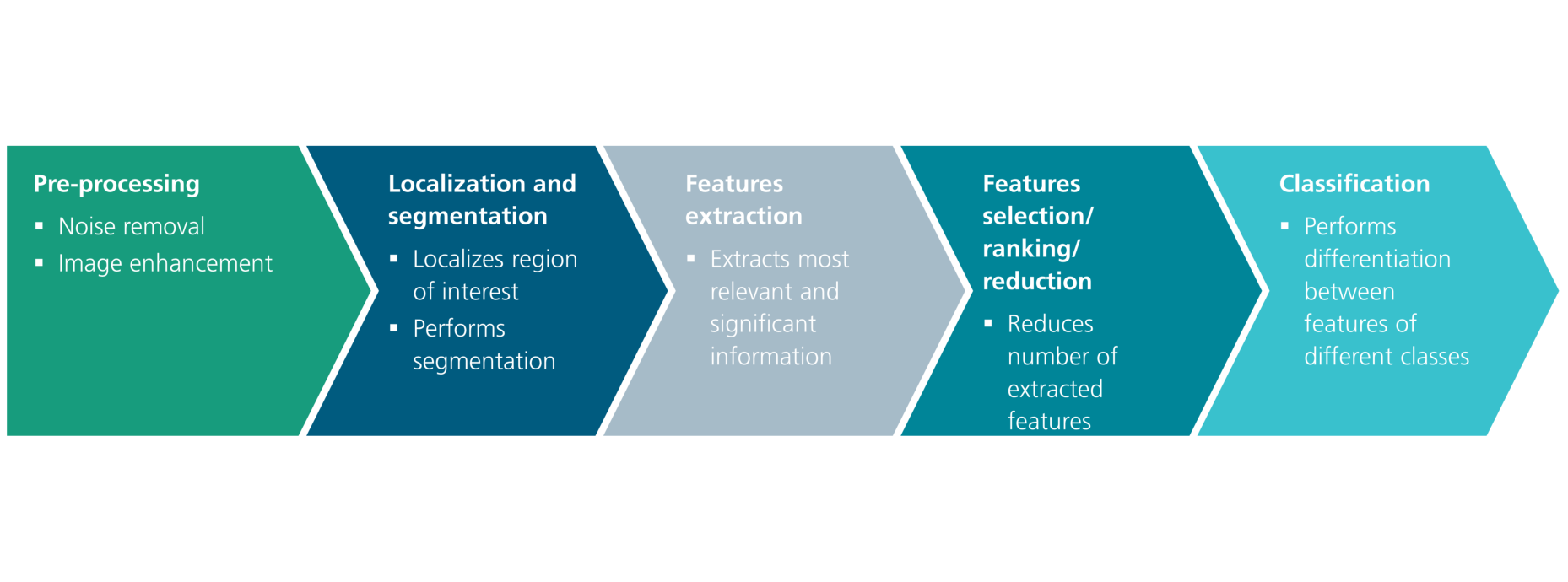}
\caption{Workflow of a typical \gls{cadx} system pipeline.}
\label{fig:workflow_ml}
\end{figure*}

With the rise in \gls{ai}, particularly \gls{dl} techniques, \gls{cadx} systems have emerged as promising tools to assist clinicians. A \gls{cadx} system is a safety-critical system within the medical field that provides automated medical diagnoses based on input medical imaging data, such as differentiating between a healthy fundus image and a glaucomatous fundus image. Figure \ref{fig:workflow_ml} depicts a typical workflow of a \gls{cadx} system which entails several key stages, including preprocessing, localization and segmentation, features extraction, features selection/reduction/ranking, and classification. 

In the development of \gls{cadx} systems, various \gls{ai} techniques play pivotal roles:
\begin{itemize}
    \item \gls{cv}: They are employed in the early stages of the \gls{cadx} pipeline, such as image preprocessing, segmentation, etc. They are often used with traditional \gls{ml} techniques. 
    \item Traditional \gls{ml}: They are often used with \gls{cv} techniques, and used mainly in the classification stage of the \gls{cadx} pipeline. 
   \item \gls{dl}: A subset of \gls{ml}, they utilize neural networks with many layers to automatically learn features from large datasets. 
    \item Hybrid approaches: Some studies combine \gls{cv}, traditional \gls{ml}, and \gls{dl} techniques to leverage the strengths of each method. 
    \item Comparison: Some studies explicitly compare the diagnostic performances of \gls{cv} and traditional \gls{ml} with \gls{dl} techniques. 
\end{itemize}

\begin{figure}[!h]
\centering
\includegraphics[width=\columnwidth]{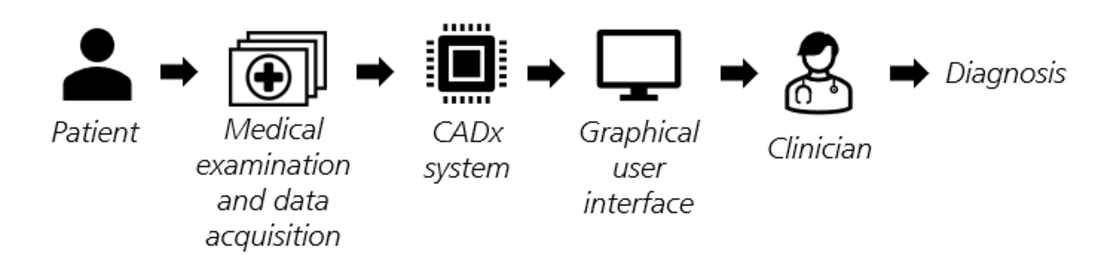}
\caption{The role of \gls{cadx} system in clinical environment.}
\label{fig:role_cadx}
\end{figure}

Regulated as medical devices, \gls{cadx} systems must undergo rigorous evaluation and compliance measures to ensure their safety and efficacy in clinical practice. Figure \ref{fig:role_cadx} describes the role of \gls{cadx} system in clinical environment. These systems can provide objective assessments, helping to standardize the diagnostic process and improve early detection accuracy if developed reliably. 

In this review, we focus on the development of \gls{cadx} systems in glaucoma diagnosis. In the context of glaucoma diagnosis, demographic information of the patient are first gathered, and then relevant clinical eye examinations are conducted. Then, eye images using various retinal imaging technologies are acquired from the patient. The data are then fed into the \gls{cadx} system for automated diagnosis. Then, the outcome of the diagnosis will be displayed on the \gls{gui} whereby the clinician can inform and provide consultation based on the output diagnosis to the patient.

The remaining paper is structured as follows. First, we provide a detailed description of our article search in Section 2, which includes the breakdown of papers into \gls{cv} and traditional \gls{ml}, \gls{dl}, hybrid techniques, and comparison papers. We include comparison papers as there are studies which specifically compare the performance of \gls{cv} and traditional \gls{ml} with \gls{dl} techniques for glaucoma detection. Then, we examine the review papers in more detail in Section 3. After that, we present our findings including the current states of commercialized systems for glaucoma detection, highlighting the research gaps and future work in Section 4. 
Finally, we conclude our paper in Section 5. 

The contributions of this paper are as follows:
\begin{itemize}
    \item We provide a comprehensive review of \gls{ai} techniques employed in glaucoma diagnosis for \gls{cadx} systems.
    \item We offer an overview of existing commercialized \gls{cadx} systems for glaucoma detection. 
    \item We highlight current research gaps in \gls{cadx} systems for glaucoma diagnosis.
    \item We propose potential future directions and methodologies to enhance the development of \gls{cadx} systems for glaucoma detection. 
\end{itemize}

\section{Search Strategy (Methodology)}
\begin{figure}[!t]
\centering
\includegraphics[clip, trim=0cm 0cm 0cm 0cm, width=\columnwidth]{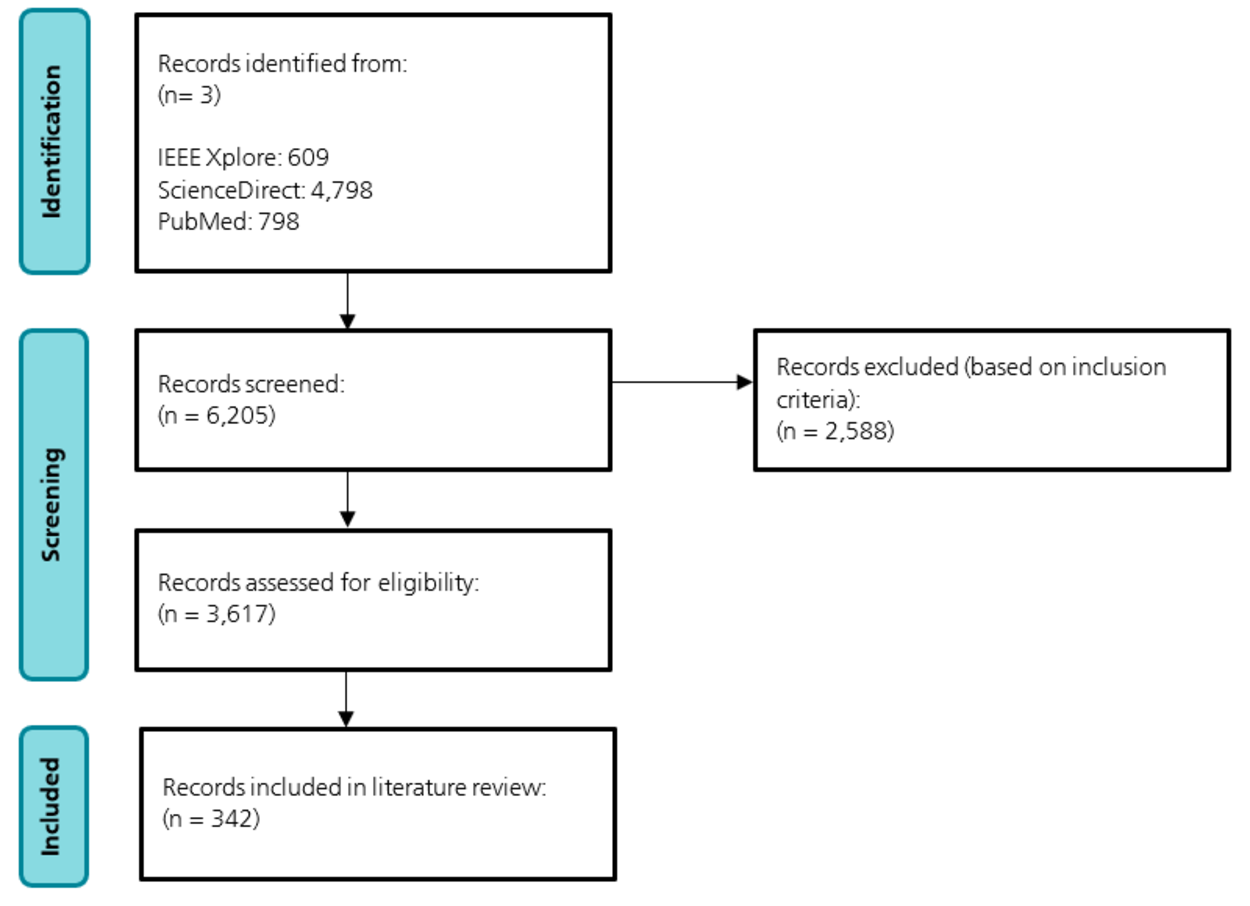}
\caption{Inclusion of relevant papers based on \gls{prisma} guidelines.}
\label{fig:prisma}
\end{figure}

A systematic search was conducted following the \gls{prisma} \cite{page_prisma_2021} guidelines to retrieve relevant studies from databases such as IEEE Xplore, ScienceDirect, and PubMed using the Boolean search string “glaucoma”, “computer-aided diagnosis system”, “medical decision support system”, “machine learning”, and “deep learning”, covering the period from 2013 to 2023 (see Table \ref{tab:bool_search_strings}). This approach aims to systematically identify relevant studies on glaucoma diagnosis using \gls{cadx} systems with \gls{ai}-based techniques. Figure \ref{fig:prisma} shows the \gls{prisma} diagram. A total of 6,205 studies were initially identified across the 3 databases. After removing duplicates, review papers, non-English publications, master’s research, and papers irrelevant to glaucoma detection using \gls{ai}-based techniques, 3,617 studies remained. Conference papers were included in this review. Then, 3,275 studies were further removed after a thorough review of abstracts and full texts, as they were deemed irrelevant to the focus of our review. The final selection yielded a total of 342 studies over the course of 10 years. 

\begin{figure}[!t]
\centering
\includegraphics[clip, trim=0cm 0cm 0cm 0cm, width=\columnwidth]{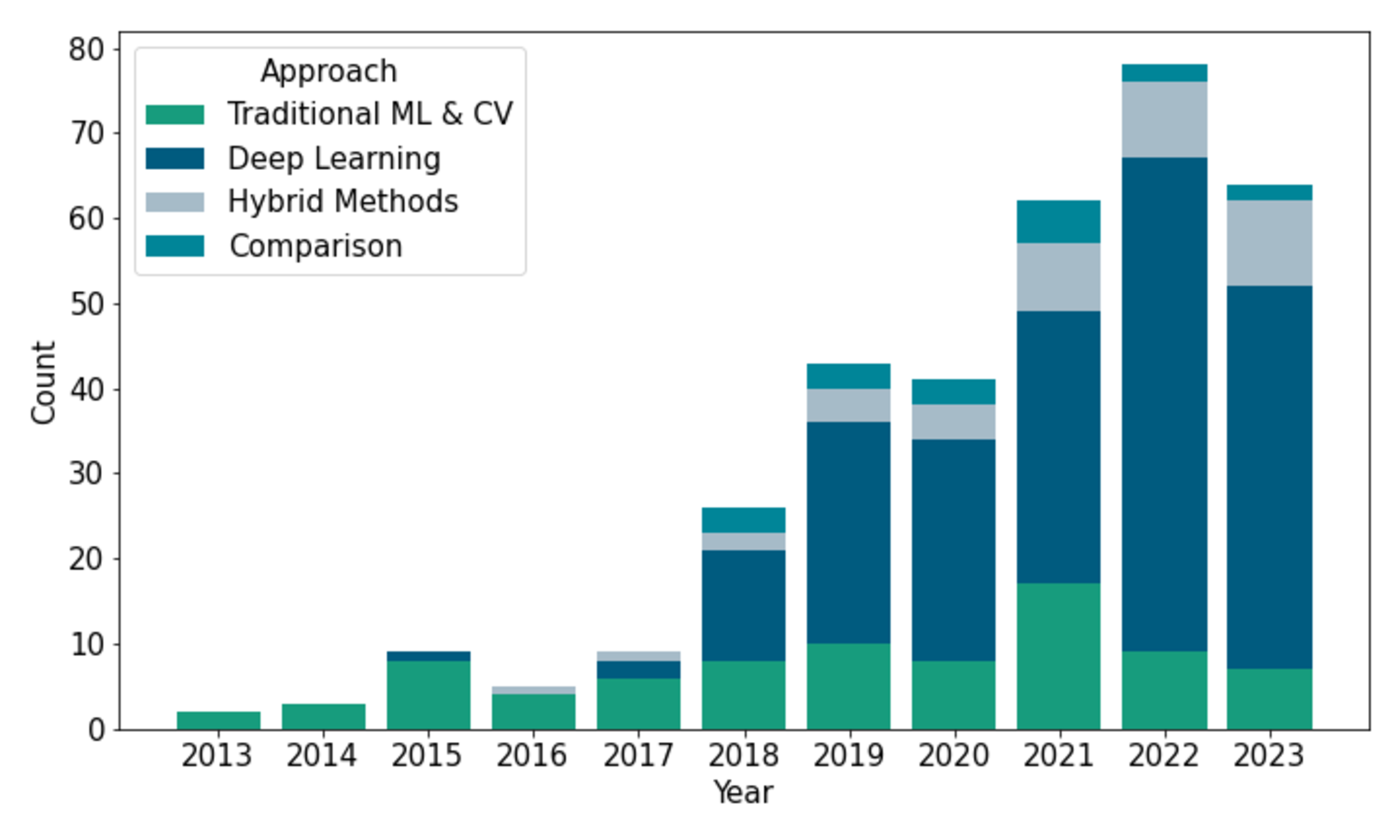}
\caption{Breakdown of the number of papers from 2013 to 2023.}
\label{fig:breakdown_num_papers}
\end{figure}

Figure \ref{fig:breakdown_num_papers} provides the breakdown of the number of papers published each year from 2013 to 2023. Generally, it can be seen that the numbers
of publications increase over the course of the years. 
The increasing number of publications each year reflects a growing interest and investment in \gls{ai}-based glaucoma diagnosis, suggesting a need for comprehensive review and analysis to identify areas for further research and development.

\begin{table}
    \centering
    \begin{tabularx}{\columnwidth}{|l|X|X|}
    \hline
        Database&  [All Metadata]& Number of studies\\
         \hline
         \rule{0pt}{20pt} IEEE Xplore&  \multirow{3}{=}{“glaucoma” AND “computer-aided diagnosis system” OR “glaucoma” AND “medical decision support system” OR “glaucoma” AND “machine learning” OR “deep learning”}& 609\\
         \cline{1-1} \cline{3-3}
         \rule{0pt}{55pt} ScienceDirect& & 4798\\
         \cline{1-1} \cline{3-3}
         \rule{0pt}{60pt} PubMed& & 798\\
         \hline
    \end{tabularx}
    \caption{\textit{Boolean search strings.}}
    \label{tab:bool_search_strings}
\end{table}

According to our literature review over the past 10 years, there have been 82 studies using \gls{cv} and traditional \gls{ml} techniques, 203 studies using \gls{dl} techniques, 39 studies employing hybrid approaches, and 18 studies specifically comparing \gls{cv} and traditional \gls{ml} with \gls{dl} techniques. 

Notably, there has been a significant surge in  \gls{dl}-based techniques over the past 6 years, coinciding with the rise in popularity of \gls{cnn} models. This upward trajectory in \gls{dl} methods has also corresponded with an increase in hybrid techniques, to integrate various \gls{ai}-based techniques to develop a more robust \gls{cadx} system. In contrast, the number of studies utilizing traditional \gls{ml}-based and \gls{cv} techniques has remained relatively constant over the years.

\section{Result}
This section provides a summary of the reviewed papers, including an overview of the data used, the medical application areas, a breakdown of the specific \gls{ml} tasks, the types of \gls{ai}-based techniques employed, and the evaluation metrics applied. 

\subsection{Data} 
In clinical practice, the diagnosis of glaucoma typically involves the integration of various types of data, including health records, tonometry to measure the intraocular pressure, clinical examination on the \gls{onh}, \gls{rnfl}, visual field testing, and advanced imaging techniques like \gls{oct}, fundus photography, and \gls{slo} \cite{european_glaucoma_society_terminology_2020}.

According to our review, a total of 281 studies utilized fundus images, 38 studies used \gls{oct} scans, 2 studies employed \gls{slo} scans, 1 study used stereo image, 1 study focused on retinal image, 2 studies utilized patient data, and 18 multiple source data, incorporating more than 1 type of imaging data or non-imaging health records.
The different types of imaging and non-imaging health records used for glaucoma detection in our review is discussed in the subsections below. 

\subsubsection{OCT data}
The \gls{oct} provides an image of structures within the retinal volume. \gls{oct} scans are able to provide measurements such as the thickness of the \gls{rnfl}, and the \gls{onh} which includes measurements like the \gls{cdr} \cite{OCT}. These clinical parameters are crucial for diagnosing glaucoma. There are a few variations of \gls{oct} scans. Spectral Domain (SD)-\gls{oct} is the advanced form of the traditional \gls{oct} while \gls{oct} angiography (OCT-A) is a functional extension of the SD-\gls{oct} by visualizing the blood flow within the retinal vasculature. It is unlike traditional angiography which requires the injection of a contrast dye, instead, it uses a motion contrast imaging to detect blood flow \cite{hagag_octa}.

\begin{figure}[!h]
\centering
\includegraphics[clip, trim=0cm 0cm 0cm 0cm, width=\columnwidth]{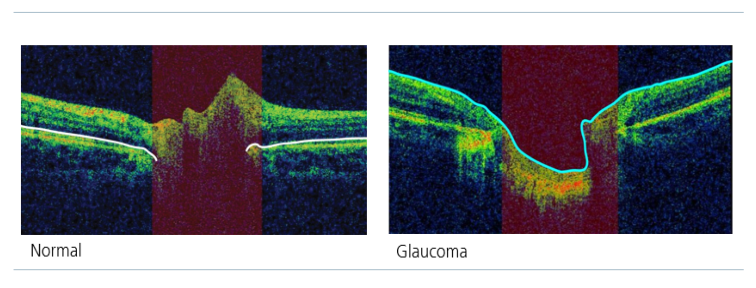}
\caption{Example of an \gls{oct} scan (taken from Mendeley dataset).}
\label{fig:oct_image}
\end{figure}

Figure \ref{fig:oct_image} depicts an example of an \gls{oct} scan. 

11 \% of the reviewed papers employed \gls{oct} data, with the majority of these studies sourcing their data privately from hospitals. There are very few publicly available \gls{oct} scans. The limited availability of publicly accessible \gls{oct} data could be a significant factor contributing to the relatively few studies utilizing \gls{oct} data.  Nevertheless, the Angle Closure Glaucoma Evaluation Challenge (AGE) \cite{AGE_oct_data} is a publicly available dataset containing \gls{oct} images. The Mendeley Dataset \cite{mendeley_oct_data} also contains labeled \gls{oct} images. Another challenge dataset, the Glaucoma \gls{oct} Analysis and Layer Segmentation (GOALS) \cite{GOALS_data_oct} dataset is also employed. 

\subsubsection{Fundus data}
On the other hand, there are many publicly available fundus datasets from various ethnicity and genders around the world. This could be the main reason why majority of these studies utilized fundus data. Approximately 82 \% of the total studies are trained with fundus images. Most of the reviewed papers employed these publicly available datasets while only a handful of papers employed their private datasets.

\begin{figure}[!h]
\centering
\includegraphics[clip, trim=0cm 0cm 0cm 0cm, width=\columnwidth]{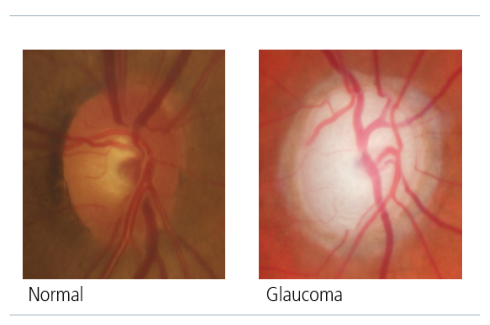}
\caption{Example of a zoomed in optic disc of a fundus image (taken from RIM-ONE DL dataset).}
\label{fig:fundus_image}
\end{figure}

The descriptions of the 10 most popular publicly available datasets employed across the literature are as follows. (1) The ACRIMA \cite{ACRIMA_2019} dataset consists of 309 healthy and 396 glaucoma images from Spain while (2) Drishti-GS \cite{drishti-gs_2014} contains 70 healthy and 31 glaucoma images from India. The Drishti-GS data also contains labels for the detection of \gls{od} and \gls{oc} for segmentation purposes. (3) The G1020 \cite{g1020_2020} also contains annotations for \gls{od} and \gls{oc} with 724 healthy and 296 glaucoma images obtained from Germany from 432 subjects. (4) The Online Retinal Fundus Image Database for Glaucoma Analysis and Research (ORIGA\textsuperscript{-light}) \cite{zhang_origa-light_2010} dataset has 482 healthy and 168 glaucoma images from Singapore with class labels also for \gls{od} and \gls{oc} and their corresponding \gls{cdr} values. (5) The PAPILA \cite{kovalyk_papila_2022} dataset from Spain consists of 333 healthy and 155 glaucoma images with \gls{od} and \gls{oc} annotated. (6) The Retinal Fundus Images for Glaucoma Analysis (RIGA) \cite{almazroa_riga_2018} dataset contains a total of 750 images from 3 sources, the Messidor dataset, Bin Rushed Ophthalmic center, and Magrabi Eye Center with only \gls{od} and \gls{oc} labels from Saudi Arabia. (7) The Large-scale Attention based Glaucoma (LAG) \cite{lag_2020} consists of 3,432 negative glaucoma and 2,392 suspicious glaucoma from China. 
(8) The Artificial Intelligence for RObust Glaucoma Screening (AIROGS) \cite{airogs_2024} is a challenge dataset that comprises a large set of over 100,000 non-referable glaucoma and nearly 5,000 referable glaucoma images from the Rotterdam EyePACS dataset. (9) The REtinal FUndus Glaucoma ChallengE (REFUGE) \cite{orlando_refuge_2020} is another challenge dataset which contains 1,080 healthy and 120 glaucoma images with a clear division between training and test images. This dataset also contains labels for \gls{od} and \gls{oc}. (10) The Retinal IMage Database for Optic Nerve Evaluation (RIM-ONE) consists of multiple versions v1, v2, v3, and DL \cite{fumero_rim-one_2011,fumero_rim_one_v2_2015,fumero_rim-one_DL_2020} obtained in different locations in Spain, 

Fundus images are high-resolution photographs which captures 30 to 50 degree views of the retina and \gls{onh}, retinal blood vessels \cite{mishra_fundus_2024}. They are crucial for the diagnosis of glaucoma. Figure \ref{fig:fundus_image} shows an example of a fundus image. 

\subsubsection{Multiple source data}
\label{sec:multi-source data}
Among the analyzed studies, 18 utilize data from multiple sources, including imaging and non-imaging data. Imaging data types from distinct sources are fundus, \gls{oct}, \gls{vf}, \gls{fa}, and \gls{lsfg}. Non-imaging data include demographics, medical history, information extracted from medical reports as well as questionnaires, which we refer to as non-imaging medical records. The most common combination of data modalities is fundus and \gls{oct}, observed in four studies. All combinations of data combinations from different sources are shown in Table \ref{tab:multiple_source_data}.

Most of these studies employ multi-modal learning techniques to directly train models using diverse data modalities as an input (section \ref{sec:multi-modal learning}). However, there are also studies that integrate multiple data sources, but ultimately apply conventional unimodal learning methods. This is done by either converting all data into numerical/categorical features, resulting in a tabular input dataset, or by using one data source as the input and another one as the output.

5 out of these 18 studies used publicly available datasets.

\begin{table*}
\centering
\caption{Overview of multiple source data used in previous studies.}
\label{tab:multiple_source_data}
\begin{tabularx}{\textwidth}{cccccccX}
\toprule
\textbf{Fundus} & \textbf{\gls{oct}} & \textbf{\gls{vf}} & \textbf{\gls{fa}} & \textbf{\gls{slo}} & \textbf{\gls{lsfg}} & \textbf{Non-imaging health records} & \textbf{Reference} \\ \midrule
x &  &  &  &  &  & x & \cite{chai_glaucoma_2018, chai_glaucoma_2021, omar_glaudia_2021} \\
x & x &  &  &  &  &  & \cite{medeiros_machine_2019,  an_glaucoma_2019, melo_ferreira_glaucoma_2023,  naqi_automated_2023} \\
x & x &  &  &  &  & x & \cite{mehta_automated_2021, lim_use_2022, li_assessing_2023} \\
x & x & x &  &  &  & x & \cite{oh_explainable_2021} \\
x &  & x &  &  &  &  & \cite{huang_detecting_2022} \\
x &  &  & x &  &  &  & \cite{hervella_2021_self} \\
x &  &  &  & x &  &  & \cite{haleem_regional_2016} \\
 & x & x &  &  &  &  & \cite{wang_towards_2020, song_asynchronous_2022} \\
 & x & x &  &  &  & x & \cite{kim_development_2017} \\
 & x &  &  &  & x & x & \cite{an_comparison_2018} \\ \bottomrule
\end{tabularx}
\end{table*}

\subsubsection{Other data}
Only a handful of 6 out of 342 papers employed other types of data in their work. There are 2 works which employed \gls{slo} scans to train their models for the detection of glaucoma \cite{abidi_data_2018, sulot_glaucoma_2021}. Similarly, the \gls{slo} provides high-resolution images of the retinal surface and allows for detailed examination of retinal structures such as the \gls{onh}, offering advantages in detecting subtle changes associated with glaucoma \cite{mohankumar_slo_2024}. 2 works utilized patient data. One of them obtained from health reports at a local hospital in India to detect glaucoma \cite{eswari_intelligent_2021}. The report consists of information such as gender, age, blood pressure, hyptertension, diabetes mellitus status, visual acuity, and \gls{cdr}. The other was obtained from Shahroud Eye Cohort Study, northeast Iran from 5,190 participants \cite{sharifi_development_2021}. They considered demographics, ophthalmologic, optometric, perimetry, and biometry features to predict glaucoma. Another paper used a stereo glaucoma image dataset in their work \cite{liu_glaucoma_2022}. Stereo images are also fundus images but they are taken as a pair with 2 cameras, resulting in 2 images (left and right) instead of 1 fundus image. Finally, 1 paper employed retinal image to train a \gls{cnn}-based model for glaucoma detection \cite{magboo_glaucoma_2023}. It is interesting to note that the retinal image employed is a publicly available \gls{ai} generated retinal image dataset available from Kaggle. 

\subsection{Medical Application Area}
Based on the literature review, we have identified various medical application areas in glaucoma diagnosis. Clinicians typically employ different medical imaging technologies (e.g., fundus photography, \gls{oct}, \gls{slo}, etc.) and non-imaging health records to diagnose a patient for glaucoma. Beyond diagnosing glaucoma, clinicians can detect the severity of the disease, identify various forms of glaucoma such as glaucomatous optic neuropathy and peripapillary atrophy detection, and differentiate between open angle and angle closure glaucoma. One common clinical parameter analyzed for glaucoma detection is the \gls{cdr} which is based on the size of the \gls{od} and \gls{oc}. Furthermore, various parameters can be extracted from \gls{oct} scans, such as vessel density measurements and \gls{rnfl} thickness, to aid in glaucoma detection. 

\subsection{Categorization of ML-specific Task}
Based on the literature review, we found that regardless of which \gls{ai}-based technique used, the studies can be broadly categorized into 3 main types: (1) Segmentation, (2) Segmentation and classification, and (3) Classification tasks. There are approximately 12.6 \%, 27.2 \%, and 58.5 \% studies in each of these categories, respectively. Within the segmentation tasks, \gls{od}, \gls{oc}, and blood vessels are commonly being localized and segmented in order to obtain the \gls{cdr} which is essential for classifying glaucoma. The \gls{rnfl} and peripapillary atrophy are also of interest to clinicians for the diagnosis of glaucoma. These features are often extracted from the images since they are clinically relevant in the diagnosis of glaucoma. Within the classification tasks, besides the prevalent binary classification (distinguishing between normal and glaucoma classes), multi-class classifications are also frequently employed. Specifically, 78.5 \% of these studies within the classification task focused on binary classification and 21.5 \% of the studies dedicated to multi-class classification, which involve either differentiating between various stages or types of glaucoma, or different eye diseases such as diabetic retinopathy and age-related macular degeneration. Additionally, a few papers, approximately 1.7 \% of the papers have included other elements such as regression, integrative index (e.g., glaucoma score, risk score, etc.), confidence scores, uncertainty scores, probabilities, and visualization techniques such as heat maps, attention maps, and saliency maps to promote explainability in their models. Nevertheless, a handful of papers focus on other aspects of \gls{ml} task such as synthesizing images, reconstructing images, domain adaptation, and out-of-distribution detection. 
Table \ref{tab:mapping_of_techniques} provides an overview of the mapping between medical application areas of glaucoma and \gls{ml}-specific main tasks. The mapping is based on a collection of review papers in this study and is by no means exhaustive. For instance, while image synthesizer is employed only in the detection of glaucoma, this does not mean that it cannot be used to generate images for the severity grading of glaucoma (i.e., normal, mild, and severe).  

\begin{table*}
\centering
\caption{Overview of mapping between medical application ares and \gls{ml}-specific tasks.}
\label{tab:mapping_of_techniques}
\begin{tabularx}{\textwidth}{llccccc}
\toprule
 & \multicolumn{1}{c}{} & \multicolumn{5}{c}{\textbf{ML-Specific Main Tasks}} \\
& \textbf{Medical Application Areas} & 
\textbf{Classification} & \textbf{Seg} & \textbf{UQ} & \textbf{Image Synthesizer} & \textbf{Visualization}  \\
\hline
 & \textbf{Detection of glaucoma} & X & X & X & X & X  \\
 & \textbf{Detection of various forms of glaucoma} & X & X &  &  &  \\
 & \textbf{Severity grading of glaucoma} & X & X & X &  & X \\
 & \textbf{Clinical parameters (e.g., CDR, RNFL)} & X & X &  &  & X \\
 \bottomrule
\end{tabularx}

\begin{flushleft}
\footnotesize{Abbreviations: Seg = Segmentation, UQ = Uncertainty Quantification, CDR = Cup-to-Disc Ratio, RNFL = Retinal Nerve Fiber Layer.}
\end{flushleft}
\end{table*}

\subsection{Computer Vision and Traditional Machine Learning Approaches} 
In accordance with traditional \gls{ml}-based and \gls{cv} techniques in the literature, the majority of studies have followed the standard workflow of a typical \gls{cadx} system (see Figure \ref{fig:workflow_ml}). 

Typically, the preprocessing, localization and segmentation, features extraction, and features selection/reduction/ranking employ \gls{cv} techniques. These steps are focused on processing and analyzing the images to extract meaningful information that can be used for diagnosis. In contrast, the classification stage employs traditional \gls{ml}-based techniques, as the classifier requires an \gls{ml} algorithm to classify images or segmented regions based on the extracted features. 

\subsubsection{Preprocessing}
Preprocessing is a crucial step performed on images before segmentation or features extraction to enhance image quality and standardize their size. Common techniques implemented for glaucoma related tasks include image standardization \cite{issac_adaptive_2015} and normalization, such as contrast stretching \cite{salam_automated_2016}, which help improve image consistency and contrast. Histogram Equalization (HE) \cite{gajbhiye_automatic_2015, raja_hybrid_2015, al-akhras_using_2021, singh_enhanced_2021}, Adaptive Histogram Equalization (AHE) \cite{noronha_automated_2014, acharya_decision_2015, acharya_novel_2017, kishore_glaucoma_2020} and its variant, the Contrast Limited Adaptive Histogram Equalization (CLAHE) \cite{maheshwari_iterative_2017, deepika_earlier_2018, panda_automated_2018, chan_automated_2019, maheshwari_automated_2019, zou_novel_2019, parashar_automated_2020, parashar_automatic_2021, shyla_automated_2021, raveenthini_combined_2021, chan_automated_2021, balasubramanian_correlation-based_2022, sonti_shape_2022, parashar_classification_2022} \cite{clahe} are widely used to enhance image quality. Additionally, various filtering techniques are employed for noise removal and texture improvement, including Histogram smoothing \cite{issac_adaptive_2015}, Median filtering \cite{deepika_earlier_2018, singh_enhanced_2021}, Wiener filtering \cite{deepika_earlier_2018, balasubramanian_correlation-based_2022}, Anisotropic filtering \cite{mohamed_automated_2019, sonti_pattern_2022}, Gabor filtering \cite{sreemol_novel_2020}, and Gaussian filtering \cite{al-akhras_using_2021, pathan_automated_2021}.   

\gls{roi} selection \cite{ruengkitpinyo_automatic_2015}, particularly focusing on \gls{od} \cite{akram_glaucoma_2015, singh_image_2016, salam_automated_2016, haleem_regional_2016}, and conversion to grayscale \cite{gajbhiye_automatic_2015, acharya_novel_2017, chan_automated_2019, thakur_classification_2020, shoba_detection_2020, elangovan_statistical_2020, chan_automated_2021, sonti_pattern_2022, sonti_shape_2022} or specific channel extraction \cite{issac_adaptive_2015, divya_performance_2018, soorya_automated_2018, jain_classification_2019, maheshwari_automated_2019, soorya_automated_2019, arab_automated_2020, al-akhras_using_2021, parashar_svm_2021, pathan_automated_2021, singh_enhanced_2021, ahmed_automated_2022} are also frequently performed to focus on relevant areas for their diagnosis tasks. 

\subsubsection{Segmentation}
Multiple studies focused specifically on the localization and segmentation of crucial clinical features such as the \gls{od}, \gls{oc}, and blood vessels to detect glaucoma. This could be due to the fact that these clinical parameters are the primary factors that help doctors with the diagnosis of glaucoma. They rely on the computation of these clinical parameters to make an appropriate diagnosis. However, it is time-consuming and highly subjective to identify these regions.
\cite{hatanaka_improved_2014} focused on the segmentation of \gls{od} and \gls{oc} by proposing an improved \gls{cv} technique using the locations of the blood vessels bends and to determine the values of the \gls{cdr} based on active contour models for the classification of fundus images into normal and glaucoma.  \cite{haleem_novel_2017} also focused on \gls{od} and \gls{oc} segmentation using Adaptive Region-based Edge Smoothing Model (ARESM) for automatic boundary detection. \cite{soorya_automated_2018} proposed seed point and bend point detection for \gls{cdr} computation following the detection of \gls{od} and \gls{oc}. \cite{mvoulana_fully_2019} employed a brightness criterion and a template matching technique to detect \gls{od} and then performed \gls{od} and \gls{oc} segmentation with a texture-based and model-based approach. Finally, they compute the \gls{cdr} values for glaucoma classification. \cite{maccormick_accurate_2019} introduced a spatial probabilistic model to derive a disc deformation index and use that as a differentiator for the detection of glaucoma. \cite{pruthi_optic_2020} proposed a Glowworm Swarm Optimization (GSO) algorithm to segment \gls{oc}. The proposed technique was demonstrated to have lesser overlapping error as compared to state-of-the-art techniques at that time. \cite{riya_active_2021} employed active contouring to detect \gls{od} and \gls{oc} boundaries with the use of filters for edge sharpening. Then, \gls{cdr} values are calculated for the classification of glaucoma.

\subsubsection{Features Extraction}
Feature extraction techniques are crucial in the analysis of medical images such as glaucoma detection to extract meaningful information from the images for diagnoses. These techniques can be grouped into various categories based on the type of features they extract. The following categories are defined based on the 83 traditional \gls{ml} and \gls{cv} review papers - clinical features, textural features, transform-based features, non-linear and entropy-based features, and statistical-based features. 
\begin{itemize}
    \item Clinical features: They are derived from specific medical measurements and anatomical properties relevant to the diagnosis of glaucoma. The common clinical features extraction based on the literature are \gls{cdr} obtained from \gls{od} and \gls{oc} \cite{ruengkitpinyo_automatic_2015, issac_adaptive_2015, akram_glaucoma_2015, salam_automated_2016, al-akhras_using_2021, kachouri_early_2023}, rim width based on the Inferior-Superior-Nasal-Temporal rule \cite{ruengkitpinyo_automatic_2015, al-akhras_using_2021}, detection of peripapillary atrophy \cite{chakrabarty_automated_2016}, retinal blood vessel \cite{singh_image_2016}, and retinal nerve fiber layer \cite{barella_glaucoma_2013, omar_glaudia_2021, fernandez_escamez_high_2021, biarnes_classifying_2023}. 
    \item Textural features: These features capture the surface properties and spatial relationships of pixel intensities in an image, providing information regarding patterns, edges, and pixel intensity distribution \cite{texture-based_features_ML}. A few commonly employed techniques found in the literature review are \gls{lbp} \cite{salam_autonomous_2015, salam_automated_2016, maheshwari_automated_2019, kishore_glaucoma_2020, ahmed_automated_2022}, \gls{lcp} \cite{acharya_novel_2017}, \gls{glcm} \cite{salam_automated_2016, cheriguene_new_2017, thakur_classification_2020, parashar_automatic_2021, paramastri_glaucoma_2021, devi_texture_2021, zulfira_segmentation_2021, chan_automated_2021, pathan_automated_2021, sonti_pattern_2022}, \gls{glrm} \cite{thakur_classification_2020, devi_texture_2021, ahmed_automated_2022}, \gls{hog} \cite{xu_automated_2013, devi_texture_2021, nawaldgi_automated_2022}, Hu moments \cite{cheriguene_new_2017, pathan_automated_2021}, and central moments \cite{cheriguene_new_2017}. 
    \item Transform-based features: These involve transforming the image into different domains to extract features. Wavelet transform \cite{salam_autonomous_2015, singh_image_2016, zou_novel_2019, thakur_classification_2020, shyla_automated_2021}, empirical wavelet transform \cite{jain_classification_2019, parashar_automated_2020}, fast Fourier transform \cite{al-akhras_using_2021}, Radon transform \cite{noronha_automated_2014, raghavendra_novel_2018, zou_novel_2019}, Gabor transform \cite{acharya_decision_2015, salam_automated_2016, sreemol_novel_2020, pathan_automated_2021, suttapakti_multi-directional_2022}, wavelet packet decomposition \cite{gajbhiye_automatic_2015}, empirical mode decomposition \cite{parashar_classification_2022}, hyper analytic wavelet transform \cite{raja_hybrid_2015}, and variational mode decomposition \cite{maheshwari_iterative_2017} are a few techniques that are being applied according to the review papers. 
    \item Non-linear and entropy-based features: These features can detect subtle pixel variations in the images. A few examples of non-linear features used in the papers are \gls{hos} \cite{thakur_classification_2020, raveenthini_combined_2021, xu_automatic_2021}, \gls{hoc} \cite{noronha_automated_2014, thakur_classification_2020}, and fractal dimension \cite{raveenthini_combined_2021}. 
    Entropy-based features capture the randomness and complexity of the images. Some examples of these entropy methods are Kapur \cite{raveenthini_combined_2021}, Yager \cite{raveenthini_combined_2021}, Shannon \cite{panda_automated_2018, raveenthini_combined_2021}, Tsallis \cite{panda_automated_2018}, and Renyi \cite{raveenthini_combined_2021}.
    \item Statistical-based features: These features extract the fundamental characteristics of the data. Some examples of statistical features are mean, variance, standard deviation, skewness, and kurtosis \cite{mohamed_automated_2019, elangovan_statistical_2020,sonti_pattern_2022}. 
\end{itemize}

\subsubsection{Features Selection, Reduction, Ranking}
The diagnostic performance of \gls{cadx} systems utilizing \gls{cv} and traditional \gls{ml}-based techniques typically depends on the features extracted. Hence, feature selection, reduction, and ranking techniques are crucial for refining the feature set to be fed into the classifier for classification \cite{feature_selection_survey}. Below are the various techniques employed based on the literature review. 

Feature selection techniques such as ReliefF \cite{parashar_2-d_2023, maheshwari_iterative_2017, thakur_classification_2020}, \gls{sffs} \cite{acharya_novel_2017} and \gls{mrmr} \cite{an_comparison_2018} are employed to identify the most relevant features. Student's t-test \cite{barella_glaucoma_2013, acharya_decision_2015, kim_development_2017, divya_performance_2018, raghavendra_novel_2018, jain_classification_2019, maheshwari_automated_2019, parashar_automated_2020, parashar_automatic_2021, chan_automated_2021}, chi-squared test \cite{barella_glaucoma_2013}, and Wilcoxon test \cite{acharya_decision_2015, raghavendra_novel_2018} are statistical methods employed to assess the significance of the extracted features in distinguishing between different classes (in this case, glaucoma or healthy). P-values are computed for each feature and they are ranked to select the most relevant and significant features for classification \cite{acharya_novel_2017}. Bhattacharyya distance \cite{acharya_decision_2015, raghavendra_novel_2018} and Fisher discriminant index \cite{noronha_automated_2014} are examples of feature ranking techniques employed to rank features according to their ability to distinguish the different classes.   

Feature reduction techniques such as \gls{pca} \cite{elshazly_chronic_2014, salam_autonomous_2015, acharya_decision_2015, singh_image_2016, salam_automated_2016, divya_performance_2018, chan_automated_2019, zou_novel_2019, parashar_svm_2021, parashar_classification_2022}, \gls{lda} \cite{noronha_automated_2014, ahmed_automated_2022}, local Fisher discriminant analysis \cite{noronha_automated_2014, akram_glaucoma_2015}, and \gls{lsda} \cite{raghavendra_novel_2018} are applied to transform and reduce the dimensionality of the feature space, enhancing computational efficiency while preserving essential information. 

To further refine the feature selection process, optimization algorithms are also employed. Algorithms such as Group Search Optimizer \cite{raja_hybrid_2015}, Particle Swarm Optimization \cite{raja_hybrid_2015}, Glowworm Swarm Optimization \cite{pruthi_optic_2020}, Gravitational Search Optimization \cite{singh_efficient_2024}, Lion Optimization algorithm \cite{balasubramanian_correlation-based_2022}, Grey Wolf Optimization \cite{balasubramanian_correlation-based_2022}, and Salp Swarm Optimization \cite{balasubramanian_correlation-based_2022} are being applied in the papers reviewed. These algorithms iteratively search for the optimal subset of features to feed into the classifier in the next step of the workflow.

\subsubsection{Classification}
The final step in a traditional \gls{cadx} pipeline is classification. Among the many classifiers, \gls{svm}, \gls{rf}, and \gls{knn} are the 3 most commonly used classifiers as per the review papers. According to these papers, there are a total of 51, 19, and 14 studies that employed \gls{svm}, \gls{rf}, and \gls{knn} classifiers respectively. Beyond these classifiers, others are also being used. A few examples are \gls{dt}, \gls{rof}, \gls{lr}, adaboost, and \gls{nb} classifiers. A handful of papers also employed ensemble classifiers \cite{barella_glaucoma_2013, kishore_glaucoma_2020, chan_automated_2021, pathan_automated_2021, sharifi_development_2021, omar_glaudia_2021, sonti_shape_2022, li_assessing_2023, singh_efficient_2024} to leverage the strengths of multiple individual classifiers and then averaging their probabilistic predictions to obtain more generalized diagnostic performances.

\subsection{Deep Learning Approaches} 
Of the 203 reviewed \gls{dl} papers, approximately 61 \% (124 papers) utilized transfer learning techniques. The remaining papers employed various methods, including \gls{cnn} (52 papers), \gls{capsnet} (2 papers), \gls{rnn} (4 papers), Autoencoders (3 papers), \gls{gan} (11 papers), and \gls{vit} (7 papers). In the subsequent subsections, we categorize and describe the reviewed papers according to these \gls{dl}-based techniques. 

\subsubsection{Convolutional Neural Network}

\gls{cnn} architectures have been widely employed for glaucoma detection due to their ability to automatically learn hierarchical features from fundus images. These architectures typically consist of multiple convolutional layers followed by fully connected layers, with techniques like dropout and data augmentation to prevent overfitting and improve robustness. Deep CNNs can be employed for classification, producing a probability distribution across various classes, or for segmentation, generating predictions for each individual pixel. \cite{chen_glaucoma_2015}, \cite{panda_deep_2018}, \cite{sharma_automatic_2019}, \cite{saxena_glaucoma_2020},\cite{manassakorn_glaunet_2022} and \cite{rakhmetulayeva_convolutional_2022} employed relatively shallow \gls{cnn} architectures comprising between one and seven convolutional layers for binary glaucoma classification, where the probability of the input image being a glaucoma is predicted. Similarly, \cite{raghavendra_deep_2018} developed an 18-layer \gls{cnn}, emphasizing the importance of removing preprocessing steps and handcrafted features to streamline the diagnostic process, while \cite{mitra_region_2018} proposed a 24-layer \gls{cnn} model to identify regions of interest in retinal fundus images. \cite{zilly_glaucoma_2017} proposed a method for segmenting retinal images using a hierarchical \gls{cnn} where filters are learned sequentially using boosting.

Established \gls{cnn} architectures such as VGG, ResNet, and others have been extensively employed to advance glaucoma detection and diagnosis. \cite{liu_development_2019} focused on binary classification problem of glaucoma in fundus images using a simplified ResNet architecture. \cite{kim_development_2020} utilized VGG-19 for glaucoma classification using thickness and deviation maps of retinal nerve fiber layer and ganglion cell–inner plexiform layer \gls{oct} images. \cite{shin_deep_2021} utilized Wide-field \gls{oct} images with VGG19-based \gls{cnn}s to diagnose glaucoma with a high degree of accuracy. Their methodology includes sophisticated fusion strategies that combine multiple \gls{oct} image views, thus capturing a comprehensive representation of the retina. \cite{juneja_deep_2022} employed a modified Xception network for glaucoma classification, while \cite{sharma_deep-glaucomanet_2023} built upon the GoogLeNet architecture, known for its efficient utilization of parameters through Inception modules, to classify glaucoma. In their work, \cite{manassakorn_glaunet_2022} introduced GlauNet, a tailored \gls{cnn} architecture designed for diagnosing glaucoma from \gls{oct}-A images, and benchmark it against VGG16, ResNet50, and EfficientNetV2. \cite{li_joint_2023} developed a region-based deep convolutional neural network (R-DCNN) based on ResNet34 for segmenting \gls{od} and \gls{oc} by treating \gls{od} and \gls{oc} segmentation as object detection problems. Notably, the network architecture included specialized components: a disc proposal network (DPN) and a cup proposal network (CPN) sequentially proposing minimal bounding boxes for \gls{od} and \gls{oc}. In a different approach, \cite{zhao_ms-ebdl_2023} focused on enhancing prediction reliability by integrating Multi-Sample Evidential Deep Learning (MS-EBDL) and multi-sample dropout into a VGG16 backbone to reduce uncertainty and improve uncertainty estimation, a critical aspect for clinical applications where prediction confidence can directly impact decision-making.

These well-established network architectures often come with pre-trained weights, which are extensively utilized in research on glaucoma detection. The goal of transfer learning is to leverage these pre-trained networks to build upon their existing knowledge, thus enhancing model accuracy and efficiency in medical imaging tasks such as glaucoma detection. They are typically trained on the ImageNet dataset \cite{deng2009imagenet}, a large-scale visual database designed for image classification and object detection tasks, containing millions of labeled images across thousands of categories. Other datasets, like Pascal VOC 2007 \cite{pascal-voc-2007}, ILSVRC (ImageNet Large Scale Visual Recognition Challenge) \cite{ILSVRC15}, and COCO (Common Objects in Context) \cite{COCO}, are sometimes also used for pre-training. Pascal VOC 2007 focuses on object detection and segmentation, ILSVRC is a subset of ImageNet, and COCO provides detailed object segmentation and recognition tasks. These networks are thereby initially trained on natural images, which may not directly resemble the medical images used for glaucoma detection. Despite this discrepancy, leveraging pre-trained weights from these datasets proves highly effective in medical applications. The networks learn to identify and abstract various features from the extensive and diverse natural images, which can be useful for detecting relevant patterns in medical images.

More recently, EfficientNet-based architectures have been used, where EfficientNet is implemented as the backbone of a \gls{cnn} model to achieve high accuracy and efficiency. \cite{yugha_automated_2022} and \cite{nawaz_efficient_2022} utilized EfficientDet-D0, with EfficientNet-B0 as the backbone, to perform precise localization and classification of \gls{od} and \gls{oc} regions. The Bi-Directional Feature Pyramid Network (BiFPN) in EfficientDet-D0 facilitates advanced feature extraction by merging information from different scales, ensuring robustness even under varying conditions such as noise and blurring. Further extending the application of EfficientNet, \cite{zhou_eards_2023} introduced the EARDS framework, which combines EfficientNet-B0 with attention mechanisms and Residual Depth-wise Separable Convolutions. This architecture segments \gls{od} and \gls{oc} regions by enhancing feature focus and computational efficiency. The integration of Attention Gates (AG) within the network ensures that relevant features are highlighted, which is crucial for the precise calculation of the \gls{cdr}, a key indicator for glaucoma.

Complementary work focuses on developing efficient models designed for glaucoma detection, aiming for deployment in resource-constrained environments. \cite{ubaidah_classification_2022} leveraged the MobileNet architecture to classify glaucoma in fundus images, emphasizing computational efficiency and speed. MobileNet's unique use of depthwise and pointwise convolutions reduces the number of parameters and operations, making it highly suitable for mobile and low-resource settings. \cite{sunija_redundancy_2022} introduced a depthwise separable convolution approach based on MobileNet to reduce computational complexity. \cite{haider_exploring_2023} introduced Efficient Shallow Segmentation Network (EE-Net) and Feature-Blending-based Shallow Segmentation (FBSS-Net), which incorporate feature blending mechanisms to improve segmentation accuracy while maintaining computational efficiency. Efficiency is crucial for real-world applications where quick, reliable diagnostics can significantly impact patient outcomes. By optimizing \gls{cnn} architectures for speed and resource use without sacrificing accuracy, these studies underscore the potential of deploying \gls{ai} in diverse healthcare environments, particularly in areas with limited access to specialized medical equipment and expertise.

Three-dimensional CNNs have been explored for glaucoma detection in notable studies. \cite{noury_deep_2022} and \cite{zang_deep-learning-aided_2023} developed such models, with Noury et al. creating a DenseNet-based architecture specifically for SD-\gls{oct} \gls{onh} cube scans, while Zang et al. employed a custom 3D \gls{cnn} architecture with 16 convolutional layers that integrates both 3D \gls{oct} and \gls{oct}-A data.

U-Net architectures and their variants have also emerged as powerful tools for the segmentation of \gls{od} and \gls{oc} in fundus images, essential for accurate glaucoma diagnosis. The foundational U-Net structure, characterized by its encoder-decoder architecture, has been adapted in several studies to address the specific challenges posed by medical image segmentation. \cite{fu_disc-aware_2018} introduced a U-Net with multi-scale inputs and side-output layers to enhance the segmentation accuracy of \gls{od} and \gls{oc}. Similarly, \cite{xu_mixed_2019} designed a U-shaped \gls{cnn} with multi-scale and multi-kernel modules to better capture \gls{oc} information of different sizes. \cite{shah_dynamic_2019} used dynamic region proposal networks based on U-Net architectures, optimizing \gls{od} and \gls{oc} segmentation through parameter sharing and dynamic cropping. \cite{shankaranarayana_fully_2019} proposed a encoder-decoder based fully convolutional network for monocular retinal depth estimation and and Optic Disc-Cup Segmentation, integrating depth maps with RGB images for enhanced segmentation performance. The work by \cite{das_cross-dataset_2022} explored the generalizability of U-Net models across different datasets. \cite{palsapure_deep_2023} and \cite{wang_automated_2021} also employed U-Net architectures, integrating in the latter various modifications such as feature detection sub-networks to preserve important image textures, thereby achieving significant improvements in segmentation accuracy. \cite{haider_artificial_2022} proposed SLS-Net and SLSR-Net, two encoder-decoder architectures inspired by U-Net aimed at minimizing spatial information loss by maintaining large feature map size for fundus image segmentation.

The use of multi-task and multi-branch \gls{cnn}s has also shown significant promise in improving the accuracy and robustness of glaucoma detection and classification. These architectures typically involve multiple stages or branches, each designed to handle different aspects of the diagnostic process. For instance, \cite{bajwa_two-stage_2019}, \cite{mojab_deep_2019} and \cite{hervella_end--end_2022} simultaneously addressed segmentation and classification tasks to enhance diagnostic accuracy. Bajwa et al. proposed a two-stage framework employing Regions with Convolutional Neural Networks (RCNN) for the localization and extraction of the \gls{od}, followed by a \gls{cnn} to classify the extracted \gls{od} as either healthy or glaucomatous, while Mojab et al. addressed the dual challenges of clinical interpretability and limited labeled data by integrating a segmentation module (U-Net) and a prediction module (VGG16). Hervella et al.'s model integrated \gls{od} and \gls{oc} segmentation with glaucoma classification, maximizing parameter sharing between tasks to leverage both pixel-level and image-level information. This method employs a multi-adaptive optimization strategy, ensuring balanced task influence during training. \cite{wang_towards_2020} integrated multiple \gls{cnn}s, including VGG16, ResNet18, ResNet50, DenseNet121, and MobileNet, in a semi-supervised multi-task learning framework to process \gls{oct} images and address missing labels. The primary task is glaucoma classification and the second one is visual field measurement regression. \cite{aljazaeri_faster_2020} also introduced a dual-step approach to localize the \gls{od}, followed by the regression of the \gls{cdr}. They combined Faster R-CNN for \gls{od} localization with DenseNet for \gls{cdr} estimation, leveraging the dense connectivity of DenseNet to enhance feature propagation and gradient flow. \cite{aamir_adoptive_2020} proposed a multi-level deep \gls{cnn} comprising detection and classification phases, which effectively identifies stages of glaucoma. Similarly, \cite{chai_glaucoma_2018} developed multi-branch networks that integrate domain knowledge with \gls{cnn}-extracted features to enhance diagnostic performance. The multi-branch approach allows these models to leverage both global and localized features, improving the overall diagnostic accuracy. Following up on their own work, \cite{chai_glaucoma_2021} then proposed a bayesian deep multisource learning model that uses Monte Carlo dropout for Bayesian inference, allowing for the estimation of predictive uncertainty, which is crucial for assessing the reliability of diagnostic results. Finally, \cite{latha_diagnosis_2022} proposed a framework for multi-disease identification, including glaucoma and diabetic retinopathy.

Previous work utilizing \gls{cnn}s has focused on self-organizing networks and unsupervised learning techniques to automatically identifying patterns and structures within data without explicit supervision. \cite{ghassabi_unified_2018} utilized a winner-take-all neural network for region of interest identification, followed by a self-organizing map for segmenting the \gls{onh} and \gls{oc}. This method effectively handles the variability in fundus images and improves segmentation accuracy. Similarly, \cite{devecioglu_real-time_2021} proposed Self-ONNs, which leverage generative neurons to self-organize during training, reducing the dependency on large annotated datasets and achieving high accuracy in glaucoma detection. In their work, \cite{zhao_diagnosing_2022} introduced a novel approach specifically tailored to address data imbalance issues in glaucoma detection. The study employs Self-Ensemble Dual-Curriculum learning (SEDC), a method that integrates self-ensembling and curriculum learning strategies.

Attention mechanisms have been effectively applied to enhance the performance of CNNs in glaucoma detection, with the aim of improving the model's ability to identify relevant features. \cite{li_attention_2019} introduced an attention-based \gls{cnn} that leverages attention maps to highlight salient regions in fundus images, thereby reducing redundant information and improving the detection of glaucoma. \cite{aljazaeri_deep_2020} utilized self-attention within an EfficientNet-B7 backbone for accurate segmentation of the \gls{oc} and \gls{od}. The use of self-attention enhances the model's ability to capture intricate details necessary for precise \gls{cdr} calculation. Similarly, \cite{dsouza_alternet-k_2024} integrated multi-head self-attention with elements from Residual Networks, demonstrating the efficacy of compact \gls{cnn} models enhanced with attention mechanisms. Furthermore, \cite{liu_interpretable_2023} built on a VGG framework with integrated attention mechanisms. Their model employs a dynamic receptive field module and a coordinate attention mechanism, allowing the model to effectively capture long-range dependencies across multiple scales. This ensures that the network focuses on the most relevant features for glaucoma diagnosis. They also employed a hierarchical gradient-weighted class activation map (CAM) function to generate activation maps that highlight the key regions within the image, making the model interpretable.

As traditional \gls{cnn}s are often regarded as "black boxes," novel architectures have been developed to provide greater transparency in their decision-making processes. For instance, \cite{keel_visualizing_2019} and \cite{kim_computer-aided_2018} displayed heat maps within regions of the \gls{onh} to visualize the predictions of their \gls{cnn} models, with Kim et al. using a VGG16-based model coupled with Grad-CAM to identify the suspicious areas. \cite{liao_clinical_2020} introduced EAMNet, an architecture designed to mimic clinical diagnostic practices by highlighting key regions in fundus images critical for diagnosis. This model utilizes Multi-Layers Average Pooling (M-LAP) with ResNet as base model to aggregate features across multiple scales, thus enhancing both diagnostic accuracy and interpretability. The distinctive approach of EAMNet, which does not depend on pixel-level annotations, simplifies its application in clinical settings by visually representing \gls{roi} that support its diagnostic conclusions. This ensures that ophthalmologists can understand and trust the model's outputs, thereby bridging the gap between automated systems and clinical practice. Such interpretability-focused advancements are crucial as they foster greater trust and reliability in \gls{ai}-driven diagnostics, addressing one of the main barriers to the widespread adoption of \gls{cnn}s in clinical environments.

\gls{capsnet} is an advanced type of neural network that enhances traditional \gls{cnn} by preserving spatial hierarchies between features. \cite{gaddipati_glaucoma_2019} and \cite{sales_dos_santos_capsule_2020} employed \gls{capsnet} using \gls{oct} and fundus images respectively. Both papers justified their use of \gls{capsnet} for glaucoma diagnosis by demonstrating that it can achieve comparable or superior performance with lesser data compared to traditional \gls{cnn} models. 


\subsubsection{Recurrent Neural Network}
\gls{rnn}s are designed to handle sequential inputs. Although \gls{rnn}s are not commonly employed in medical imaging applications, there are a few studies that utilize \gls{rnn}s for glaucoma detection. 

Incorporating recurrent components for the glaucoma detection task, \cite{garcia_glaucoma_2021} presented a novel combined \gls{cnn}-\gls{lstm} architecture. While the end-to-end solution is based on 3D spectral domain \gls{oct} data, the convolutional slide-level feature extractor is pre-trained on circumpapillary data. Wherefore, the authors acknowledged differences in patterns but note the similarity in the involved structure. The convolutional, VGG16-based component consists of residual and attention modules. 
Considering the spatial dependencies as a temporary instance, the \gls{lstm} is then fed with the extracted features from one slide per step. To integrate \gls{lstm} outcomes at different steps for the classifier, the authors developed a sequential weighting module.\cite{gheisari_combined_2021} relied on \gls{lstm} in a slightly different use case to improve glaucoma detection based on image and video fundus data. By leveraging the temporal sequence of the video data and spatial feature extraction for each frame with VGG16 and ResNet50, the authors reported substantial performance improvements over using VGG16 or ResNet50 alone. \cite{kumar_novel_2023}, preprocessed retinal fundus images using HE technique and CLAHE, after which a U-Shape network, namely Unet++, segmented \gls{od} and \gls{oc} image portions. For each of these segments, the authors used their proposed Residual Network with Gated Recurrent Units for glaucoma disease detection (separately for disease stage 1 and 2), which were finally combined for overall detection. \cite{raja_damped_2019} proposed a Damped Least-Squares Recurrent Deep Neural Learning based on fundus image data. For effective glaucoma disease detection, the authors relied on recurrent feedback connections. In this context, the outputs of the prior hidden layers are fed to the input layer accompanying current fundus image inputs.

\subsubsection{Autoencoders}
The autoencoder is an unsupervised representation learning technique that relies on an encoder decoder network architecture to (1) transform the input data into a lower-dimensional encoded representation and (2) reconstruct the original input data from this encoded representation. This approach is particularly useful in medical applications, such as glaucoma detection, where large publicly available datasets are limited.  

\cite{pal_g-eyenet_2018} proposed a G-EyeNet which combines a deep convolutional autoencoder and a traditional \gls{cnn} classifier. It first segments the \gls{od} via U-Net then fed the segmented image into their proposed model. They used autoencoders in their work as they mentioned that autoencoders learn a low-dimensional representation of an input and therefore, more suitable to learn priors on the distribution of data. They reported that their proposed G-EyeNet model has outperformed two other techniques, a state-of-the-art \gls{cnn} model and a hybrid of \gls{cnn} and \gls{svm}. 
\cite{raghavendra_two_2019} proposed to cascade two sparse autoencoders with a softmax layer to classify normal and glaucoma images. Their proposed technique outperformed various traditional \gls{ml} classifiers and \gls{cnn} models. \cite{al_ghamdi_semi-supervised_2019} employed a semi-supervised 6-layer \gls{cnn} model with denoising autoencoder and compared it with a simple transfer \gls{cnn} model and semi-supervised \gls{cnn} model with self-learning. They demonstrated that the model with autoencoder achieved the highest diagnostic performance as compared to the other two techniques. 

\subsubsection{Generative Adversarial Network}
\gls{gan}, with their ability to generate synthetic data, improve image resolution, and adapt to different imaging conditions, have been extensively researched for their potential to improve glaucoma detection.

The utilization of \gls{gan}s to perform image synthesis and augment data for glaucoma classification first addresses the issue of data scarcity, enhancing the diversity and quality of training datasets. \cite{diaz-pinto_retinal_2019} employed Deep Convolutional GANs (DCGANs) to create synthetic images that improve the performance of semi-supervised learning models, achieving high classification accuracy by leveraging a small labeled dataset combined with a large unlabeled one. The DCGAN architecture consists of a generator and a discriminator, where the generator creates realistic synthetic images and the discriminator distinguishes between real and synthetic images. This approach utilizes a binary cross-entropy loss for the discriminator and a combination of adversarial and reconstruction losses for the generator to ensure high-quality image synthesis. Similarly, \cite{leonardo_impact_2022} utilized CycleGANs to transform fundus images, focusing on enhancing and degrading image quality to improve diagnostic models. This approach emphasizes the role of high-quality image diversity in training more effective classifiers. \cite{manikandan_glaucoma_2023} introduced CAPSGAN, a hybrid model integrating \gls{gan}s and \gls{capsnet} to produce synthetic images that augment the dataset, highlighting the importance of preserving spatial and orientational details for early glaucoma detection. 

\gls{gan}s have also shown promise in improving the segmentation of critical features like the \gls{od} and \gls{oc}. \cite{wang_patch-based_2019} employed a novel patch-based output space adversarial learning framework to segment the \gls{od} and \gls{oc} from various sources of fundus images. The architecture includes a lightweight and efficient segmentation network as the backbone, featuring a morphology-aware segmentation loss and unsupervised domain adaptation. \cite{bian_optic_2020} leveraged conditional \gls{gan}s (cGANs) to achieve precise segmentation by incorporating anatomical guidance, mimicking the manual examination process. The c\gls{gan} architecture used in this study includes a modified U-Net as the generator for mask generation and a discriminator based on LeNet. The loss function combines adversarial loss and pixel-wise binary cross-entropy loss to ensure accurate segmentation. This method significantly enhanced segmentation accuracy by utilizing a cascade network focused on small target areas. \cite{pachade_nenet_2021} further improved segmentation accuracy through the Nested EfficientNet (NENet),  which integrates adversarial learning with EfficientNetB4 as the encoder, incorporating pre-activated residual blocks, atrous spatial pyramid pooling (ASPP), and attention gates (AGs). The network is trained using a combined cross-entropy and dice coefficient loss to improve segmentation precision, and a modified patch-based discriminator to enhance local segmentation details. The use of SRGAN by \cite{nandhini_detection_2022} for super-resolution image enhancement demonstrates how higher resolution images can improve segmentation outcomes and subsequent classification accuracy. 

Lastly, \gls{gan}s have been pivotal in addressing domain adaptation challenges in glaucoma diagnosis, ensuring that models perform robustly across different imaging devices and datasets. \cite{sun_gan-based_2020} identified a significant issue in medical image analysis where variations in imaging devices lead to domain shifts that negatively impact the performance of diagnostic models. They introduced a domain adaptation method for glaucoma diagnosis, leveraging Wasserstein \gls{gan} with Gradient Penalty (WGAN-GP) loss for enhanced stability in domain adaptation, significantly improving model performance across different datasets. WGAN-GP uses a gradient penalty term to improve training stability. \cite{chourasia_domain_2023} utilized deep convolutional GANs (DCGAN) to reduce domain shifts, demonstrating robust diagnostic performance by generating target-like source images. The DCGAN architecture includes convolutional layers for both the generator and discriminator. \cite{liu_ecsd-net_2022} proposed ECSD-Net, which employs \gls{gan}s for unsupervised domain adaptation to enhance segmentation and classification accuracy, addressing the distribution gaps between datasets. ECSD-Net integrates EfficientNet for feature extraction, and a bi-directional feature pyramid network for feature fusion. The loss function combines adversarial loss, segmentation loss, and classification loss to improve overall performance.\cite{zhou_unsupervised_2023} introduced MAAL, a Multi-scale Adaptive Adversarial Learning framework that effectively mitigates model degradation due to domain shifts. The MAAL framework includes a Multi-scale Wasserstein Patch Discriminator (MWPD) designed to extract domain-specific features, an Adaptive Weighted Domain Constraint (AWDC) to further improve model generalization, and a Pixel-level Feature Enhancement (PFE) module to enhance low-level features obtained from shallow layers.

\subsubsection{Vision Transformers}
\gls{vit}s utilize a self-attention mechanism to process image data, enabling the model to capture complex patterns and relationships across different parts of the image. \gls{vit}s are an emerging technique that is gradually gaining recognition in medical applications. Recent studies have shown that \gls{vit}s can surpass the diagnostic performance of traditional \gls{cnn} models. 

\cite{wang_workflow_2022} employed various trained models in their proposed workflow for the classification of referrable, non-referrable, and ungradable glaucoma. First, they used ResNet101-upernet as the backbone model for \gls{od} segmentation. Then, they utilized U-Net for retinal vessel segmentation. After which, they applied \gls{vit} to classify referrable and non-referrable glaucoma. They also developed a model with ResNet18 to detect ungradable glaucoma based on the segmentation of \gls{od} and retinal vessel. Their proposed technique achieved first place in the AIROGS challenge in 2022. \cite{riza_rizky_adversarial_2022} focused on developing a robust system to detect the severity of glaucoma. They trained 3 models including \gls{vit}, ResNet50 model, and Gated Multilayer Perceptron (gMLP) with adversarial training and deep \gls{knn}. They have demonstrated that adversarial training with Deep \gls{knn} can improve the model’s robustness. \cite{wassel_vision_2022} proposed an ensemble of transformer-based models including DeiT, BEiT, CaiT, ResMLP, XCiT, CrossViT, and Swin with transfer learning to detect glaucoma using fundus images. \cite{xue_cts-net_2023} proposed a method for retinal layer segmentation based on CSWin Transformer (CTS-Net). They also employed a dice loss function based on boundary areas in order for the proposed model to learn more features along the edges and to improve segmentation accuracy. \cite{fan_detecting_2023} employed Data Efficient Image Transformer (DeiT) model to detect glaucoma from fundus images and compared it to the state-of-the-art ResNet50 model. Their results demonstrated that the DeiT model outperformed ResNet50 model when validated with external dataset. \cite{haouli_exploring_2023} proposed a \gls{vit} based method for the detection of glaucoma. In order to demonstrate that \gls{vit} are better than \gls{cnn} models, they compared their proposed model with Xception and ResNet152V2 models. The results have shown that \gls{vit}-based models achieve better diagnostic performances compared to \gls{cnn}-based models. \cite{hwang_multi-dataset_2023} aim to compare the diagnostic accuracy of \gls{vit}-based and \gls{cnn}-based models in the detection of glaucoma using fundus images. They have demonstrated that \gls{vit}-based models often show superior performances particularly when the dataset is imbalanced. All in all, these authors have a common goal of enhancing glaucoma detection through employing \gls{vit} algorithms. By leveraging on \gls{vit}-based algorithms, they seek to overcome the drawbacks inherent in \gls{cnn} algorithms, and to improve the robustness and generalizability of the \gls{cadx} systems.

\subsection{Hybrid Approaches} 
As per the literature, hybrid techniques combine both \gls{cv} and traditional \gls{ml} with \gls{dl} techniques to enhance diagnostic accuracy. A total of 39 papers utilized these hybrid techniques to develop more effective \gls{cadx} systems. These 39 studies can be further categorized based on the specific types of hybrid approaches used. 

A minority of the studies relied on \gls{dl} for \gls{od} and \gls{oc} segmentation and used clinical or \gls{cv}-based features for final classification. For instance, \cite{tulsani_automated_2021} developed a custom UNET++ approach for binary \gls{oc} and \gls{od} masks that are input to clinical features used for classification. \cite{veena_novel_2022} employed two separate \gls{cnn}s for \gls{od} and \gls{oc} semantic segmentation using preprocessed and extracted inputs inferred via \gls{cv} methods, which are further used for the \gls{cdr} to inform about the glaucoma progression. Similarly, \cite{lee_screening_2019} used a pre-trained \gls{dl} approach for glaucoma classification and Polar Transformation for the localization of \gls{rnfl} defects. 

From our literature review, it can be inferred that combining segmentation and classification tasks or focusing on classification alone is more common than focusing solely on segmentation task. Regardless of the \gls{ml}-based application task, the majority of these studies leveraged on transfer learning by using pre-trained models.

Several studies employed pre-trained models for automated features extraction followed by traditional \gls{ml} classifiers for glaucoma detection. For instance, \cite{al-bander_automated_2017,balasubramanian_improved_2022,kamini_analysis_2023} used \gls{svm} classifier in their work whereas \cite{de_moura_lima_glaucoma_2018} and \cite{gunasinghe_comparison_2021} employed pre-trained models as feature extractors then fed into various traditional \gls{ml} classifiers including \gls{lr}, \gls{svm}, Naive Bayes, and \gls{rf}. Furthermore, \cite{sharma_automated_2023} used an ensemble of pre-trained models to automatically extract relevant features and feed them into the modified least-squares \gls{svm} classifier for glaucoma detection. \cite{li_integrating_2016} leveraged pre-trained \gls{cnn}s fed with full and patched image inputs for holistic and local feature extraction, respectively. Based on these features the authors used separate \gls{svm} classifiers and derived the final classification results using decision fusion.

\cite{zhao_direct_2020} presented their \gls{cnn}-based MFPPNet for immediate \gls{cdr} value prediction without intermediate segmentation, using the pre-trained \gls{dl} component for unsupervised feature extraction and the \gls{rf} classifier to derive the final \gls{cdr} value.

On the other hand, rather than using pre-trained models, \cite{chakrabarty_novel_2019} employed a \gls{cnn} model for feature extraction, combined with a \gls{svm} classifier. 
In a related work, \cite{bouacheria_automatic_2020} extracted the \gls{roi} using a \gls{cnn}. Afterward, the authors relied on \gls{cv} techniques for \gls{oc} \gls{od} segmentation, which was the input for the \gls{cdr}, \gls{isnt} rule, and \gls{ddls} features, which were the input to a \gls{rf} classifier.

Some studies combined pre-trained models with \gls{cv} techniques for feature extraction followed by traditional \gls{ml}-based classifiers for glaucoma detection.
For instance, \cite{claro_hybrid_2019} extracted more than 30K features from segmented images using \gls{cv} methods and seven pre-trained \gls{cnn}s. Subsequently, the authors leveraged the gain ratio algorithm to select the most important features for input to a \gls{rf} classifier.

Some of them used \gls{cv} techniques for feature extraction with pre-trained models for classification.
\cite{chaabane_glaucoma_2023} compared several pre-trained \gls{dl} models for classification with transfer learning and leveraged Gabor filters and \gls{pca} for texture extraction.
\cite{chaudhary_automatic_2022} derived ResNet-50 from sub-images generated by \gls{2d_fbse_ewt}. The authors reduced the concatenated per-sub-image-features through \gls{pca} for a final classification with a softmax classifier to distinguish between \gls{poag}, \gls{pacg}, secondary glaucoma, healthy.  

On the other hand, one study employed \gls{cv} and traditional \gls{ml} methods for segmentation and \gls{dl} techniques for classification. For example, \cite{singh_optimized_2023} used Modified Level Set Algorithm for \gls{oc} segmentation and classification via optimized \gls{cnn}, based on extracted morphological and non-morphological features. 

\cite{vinicius_dos_santos_ferreira_convolutional_2018} combined \gls{dl} for segmentation with phylogenetic diversity indexes feature extraction followed by a \gls{cnn}-based model for glaucoma classification.

\cite{fang_glaucoma_2023} presented the \gls{gsm_dcn}, which combines features derived from clinical knowledge, anatomical feature maps, and the glaucoma syndrome mechanism as input for multi-classification (normal, mild, moderate, severe).

Other methods relied on a combination of conventional \gls{ml} and \gls{dl} approaches for glaucoma classification.

For example, \cite{charulatha_robust_2022} used \gls{cv} methods for \gls{roi} extraction, while (partially pre-trained) \gls{dl} approaches were used for input verification, segmentation, and classification, which was also supported by the Light GBM, traditional \gls{ml}-based method.
After preprocessing and \gls{roi} extraction, \cite{ko_deep_2020} relied on transfer learning based on VGGNet for classification. Notably, the authors complemented the classification with a \gls{svm} classifier based on the \gls{isnt} rule and \gls{cdr} for low-confidence cases.

Moreover, from our literature, it is observed that all the studies employed U-Net or variant of U-Net architectures for segmentation combined with either traditional \gls{ml} classifier or \gls{dl} approaches for classification \cite{shanmugan_automatic_2021, shinde_glaucoma_2021, bajaj_fundus_2023, singh_self-improved_2023, naqi_automated_2023, sangeethaa_presumptive_2023, ganapathy_evaluation_2022}. 

While the majority of these studies used \gls{cnn}-based architectures, a handful employed other \gls{dl} techniques, including \gls{gan}, \gls{capsnet}, and autoencoder. For example, \cite{bisneto_generative_2020} applied \gls{gan} to efficiently segment the \gls{od} and then used the taxonomic diversity index to extract features from the segmented \gls{od} for automated glaucoma detection. Similarly, \cite{jain_rider_2022} introduced a novel optimization-based \gls{gan} integrated with \gls{cnn} features, focusing on early detection through enhanced segmentation and detailed feature extraction using various \gls{cv} techniques. 
In another study, \cite{singh_novel_2022} developed a hybrid classification fusion model combining traditional \gls{ml} classifiers with DCGAN and VGG-\gls{capsnet} architecture. Additionally, \cite{sanchez-morales_improving_2022} leveraged 3 \gls{dl}-based approaches, \gls{cnn}, \gls{capsnet}, and autoencoder architectures and combined their outputs with a \gls{knn} classifier, demonstrating that ensemble learning methods can improve state-of-the-art results in glaucoma detection.  

Further, reinforcement learning has also been implemented for glaucoma detection. For instance, \cite{naseeha_abdulla_precise_2023} proposed a deep reinforcement learning model with a deep Q network for active object localization to detect the \gls{od} region. They combined traditional \gls{ml} and \gls{cv} methods with \gls{cnn} for segmentation and feature extraction before applying reinforcement learning.

Other novel methodologies within hybrid approaches include the works by \cite{girard_atlas-based_2021} and \cite{xu_hierarchical_2021}. The former proposed an atlas glaucoma score to detect abnormal deformations induced by glaucoma, while the latter developed a hierarchical deep learning system which includes 3 modules: pre-diagnosis, segmentation, and final diagnosis based on expert knowledge for glaucoma diagnosis.  

Additionally, 4 studies \cite{mehta_automated_2021, lim_use_2022,oh_explainable_2021,an_glaucoma_2019} utilized multi-modal learning in their hybrid approaches. The details are described in \ref{sec:multi-modal learning}. 

\subsection{Comparison Papers of Traditional \gls{ml} and \gls{cv} with \gls{dl} Techniques}
18 studies have employed different techniques to evaluate and compare the performance of \gls{dl} with \gls{cv} and traditional \gls{ml} models in the diagnosis of glaucoma. \cite{touahri_comparative_2018} employed \gls{cnn} models with 2 and 3 convolution layers and compared them with traditional \gls{ml} approaches using central moments, Hu moments, and GLCM extraction techniques with a twin \gls{svm} classifier. \cite{ajitha_identification_2021} compared a 13-layer \gls{cnn} model with an \gls{svm} classifier and a \gls{cnn} model with softmax layer, showing that combining \gls{cnn} and \gls{svm} produced superior diagnostic performance. \cite{akter_glaucoma_2022} compared the diagnostic ability of a 24-layer \gls{cnn} model and \gls{lr} classifier.

Out of the 18 studies, 9 studies have highlighted the efficacy of transfer learning and pre-trained models in enhancing the performance of \gls{dl} techniques, often showing superior performance compared to traditional \gls{ml} and \gls{cv} approaches \cite{asaoka_using_2019, pandey_detection_2020, zheng_detecting_2020, ahn_deep_2018, shibata_development_2018, thakoor_enhancing_2019, sulot_glaucoma_2021, sunanthini_comparison_2022, abraham_comparative_2023}.

\cite{maetschke_feature_2019} compared the performance of traditional \gls{ml} using 22 clinical features and feeding them into various classifiers with a 3D \gls{cnn} architecture. Similarly, \cite{george_attention-guided_2020} utilized the same 22 clinical features with 6 traditional \gls{ml}-based classifiers, comparing them against their proposed end-to-end guided \gls{dl} framework. \cite{chaudhary_automatic_2021} evaluated the efficacy of their proposed 2D-Fourier-Bessel Series Expansion based Empirical Wavelet Transform (2D-FBSE-EWT) using 2 approaches: (1) a traditional \gls{ml} and \gls{cv} technique with \gls{glcm} and moment features extraction, followed by feature reduction and various feature ranking techniques before feeding these features into \gls{mlp}, \gls{svm}, least squares \gls{svm}, and \gls{rf} classifiers and (2) an ensemble of ResNet50 model combined with the different features derived from 2D-FBSE-EWT. The results demonstrated that both traditional \gls{ml} and \gls{cv} and \gls{dl} models performed comparably well. \cite{xu_automatic_2021} proposed a Transfer-Induced Attention Network (TIA-Net) and compared it with (1) a \gls{lr} classifier fed with \gls{hos}, discrete wavelet transform, Gabor transform features, and (2) classical transfer learning \gls{cnn} models such as VGG, GoogLeNet, and ResNet, demonstrating that their proposed model outperformed both traditional \gls{ml}, \gls{cv}, and classical \gls{cnn} techniques. Lastly, \cite{suganthe_glaucoma_2023} compared the performance of a \gls{cnn} model against \gls{rf} and \gls{svm} classifiers in glaucoma diagnosis and showed that the \gls{cnn} model outperformed both \gls{rf} and \gls{svm} classifiers. Conversely, \cite{zulfira_segmentation_2021} proposed using a traditional \gls{ml} technique, a dynamic ensemble selection classifier to detect glaucoma severity and compared this approach with the U-Net model. The results showed that the traditional \gls{ml} technique outperformed the U-Net model, likely due to the U-Net's limited learning capability with small datasets and the absence of pre-training. 

Based on these 18 papers, almost all the papers have reported \gls{dl} techniques to outperform traditional \gls{ml} and \gls{cv} approaches in terms of diagnostic accuracy and other performance metrics. Furthermore, some studies have shown that the use of state-of-the-art \gls{dl} techniques have superior or even better performance than radiologists in some diagnostic tasks~\cite{Chan2020}.

\subsubsection{Multi-modal learning}
\label{sec:multi-modal learning}
Multi-modal learning refers to the integration of data of different modalities from various sources. In glaucoma research, multi-modal data can include different types of imaging data (e.g., fundus image, \gls{oct}, \gls{slo} as well as numerical/categorical features demographics, clinical records, \gls{vf} data). When developing \gls{ai}-based systems, utilizing multiple data modalities allows for integrating complementary information which can enhance model performance and improve prediction robustness. At the same time, this comes at the cost of higher problem complexity, requiring more advanced approaches able to leverage these modalities and handle data heterogeneity. 

Among the analyzed studies, 11 employ multi-modal learning techniques. 10 out of these utilize modalities from varying sources, such as fundus image and numerical/categorical patient data (section \ref{sec:multi-source data}). 1 study, however, utilizes data from a single source, specifically fundus images, and additionally extracts numerical features such as \gls{cdr} and eccentricity values, ultimately using both data modalities as input \cite{das_cross-dataset_2022}.

Within the group of 10 studies that integrate data from multiple sources, various methodologies are employed. 2 studies utilized \gls{cnn} approach. Chai et al. employed  fundus images and non-imaging health records to develop \gls{cnn}-based models for glaucoma detection in both their 2018 study \cite{chai_glaucoma_2018} and in their subsequent 2021 work \cite{chai_glaucoma_2021}.  

4 studies applied transfer learning techniques, leveraging pre-trained models to enhance the performance. For instance, \cite{hervella_2021_self} proposed a self-supervised visual learning strategy and multimodal reconstruction with fundus and \gls{fa} for the diagnosis of age-related macular degeneration and glaucoma. \cite{huang_detecting_2022} employed a probabilistic Bayesian model with EfficientNet4 architecture that provides certainty scores using fundus images and \gls{vf} data, aiming to give clinicians greater confidence in the model's diagnoses. \cite{melo_ferreira_glaucoma_2023} used fundus images and 3D \gls{oct} volumes to develop a \gls{dl}-based model using various pre-trained models to grade the different stages of glaucoma. \cite{li_transfer_2023} developed a novel multimodal neural network for glaucoma diagnosis and classification (GMNNet) with fundus images, \gls{oct} scans, and non-imaging health records.   

Moreover, 4 studies adopted a hybrid approach, combining \gls{cv} and traditional \gls{ml} with \gls{dl} techniques. For example, \cite{an_glaucoma_2019} combined fundus images and 3D \gls{oct} data to develop a model using VGG19 architecture and \gls{rf} classifier for glaucoma diagnosis. Similarly, \cite{oh_explainable_2021} introduced a glaucoma prediction model with an explanation function, incoporating fundus images, \gls{oct} scans, \gls{vf}, and non-imaging health records. \cite{mehta_automated_2021} built separate models using fundus images, \gls{oct} scans, and non-imaging health records, then combined the results to construct a final model with gradient-boosted \gls{dt} for glaucoma detection. Likewise, \cite{lim_use_2022} utilized fundus images, \gls{oct} scans, and non-imaging health records to develop a model that combined the \gls{dl}-based Xception architecture for feature extraction with traditional \gls{ml} classifiers, categorizing cases into normal, pre-perimetric glaucoma, and glaucoma classes. 

Nevertheless, 1 study employed cross-modal learning. It involves transferring knowledge or information between different modalities. \cite{song_asynchronous_2022} developed a framework combining asynchronous feature regularization and cross-modal knowledge distillation from \gls{vf} and \gls{oct} networks for enhancing glaucoma diagnostic performance on \gls{oct} modality.

\subsection{Evaluation Metrics} 
Drawing from the literature, several common metrics are employed to evaluate the performance of \gls{cadx} systems in glaucoma. Among these metrics, \gls{acc}, \gls{sen}, and \gls{spe} are frequently used to assess the overall performances of \gls{cadx} systems in correctly identifying and classifying glaucoma using various types of data. However, in medical applications, including glaucoma detection, datasets are often imbalanced with a disproportionate number of diseased compared to healthy ones. In such scenarios, accuracy metric alone may be misleading, as the \gls{cadx} system could achieve high accuracy by simply predicting the majority class. 

To address this, additional metrics like the \gls{auc} and the \gls{auroc} are utilized to measure the discriminative ability of the models. A handful of studies also employed \gls{f1} and Precision metrics, which are particularly useful for evaluating the performance in the presence of imbalanced data, as they provide a better understanding of how well the \gls{cadx} system detects glaucoma, considering both \gls{fp} and \gls{fn}. 

Furthermore, the dice similarity coefficient is often employed to quantify the spatial overlap between the segmentation results of the \gls{od}, \gls{oc}, and blood vessels and the ground truth annotations. Overlapping error and intersection over union metrics are also utilized to quantify the degree of overlap between segmented regions and ground truth annotations especially in the segmentation of \gls{od} and \gls{oc}. These metrics offer valuable information on segmentation accuracy and boundary delineation performance. 

\section{Discussion}
This section discusses the current state of commercialized \gls{cadx} systems for glaucoma detection and then identifies current research gaps in this field.  

\subsection{Commercialized CADx Systems for Glaucoma Detection}
The various available commercialized \gls{cadx} systems for glaucoma detection are: 
\begin{itemize}
    \item RETeval by LKC Technologies \cite{reteval}: measures ganglion cell function or optic nerve disorders to prevent further degeneration of ganglion cells that is associated with glaucoma disease using electroretinogram (ERG) technology. Unlike typical imaging technology, the ERG measures the electrical responses of various cell types in the retina. 
    \item RetinaScope Plus by Visulytix \cite{retinascope}: a handheld, non-mydriatic retinal imaging camera for retina screening of eye diseases including diabetic retinopathy, age-related macular degeneration, and glaucoma using a RS+ software. 
    \item iCare ILLUME \cite{icare}: captures high-quality images and uses \gls{ai}-based technology for the detection of early signs of diabetic retinopathy, age-related macular degeneration, and glaucoma. It returns a glaucoma score which indicates the likelihood for glaucoma together with useful heatmaps that indicate areas with possible abnormalities. 
   \item RetinaLyze \cite{retinalyze}: uses \gls{ai}-based technology for the screening of multiple eye diseases using fundus images and \gls{oct} scans. A glaucoma discriminant function using the colours on the fundus images is used to determine the damage to the \gls{onh}, which serves as an indication of the presence of glaucoma. 
\end{itemize}
 
While several commercialized systems are available, the number specifically designed for glaucoma detection is limited compared to those for other eye diseases, such as age-related macular degeneration and diabetic retinopathy. This disparity may be due to several factors, including the complex nature of the disease, the absence of well-defined diagnostic criteria, the asymptomatic progression of the disease in its early stages, and the variability of symptoms across different patients \cite{glaucoma_risk_factors}.

The limited number of commercialized \gls{cadx} systems specifically for glaucoma highlights a significant gap in the market and underscores the need for more focused research and development in this area. 

\subsection{Current Research Gaps} 
Our literature review reveals that contributions concerning the reliability of \gls{cadx} systems are notably sparse. Despite the literature review showcasing a plethora of novel proposed methodologies and state-of-the-art \gls{ai}-based techniques in glaucoma diagnosis, only a fraction of the literature, less than 10 \%, encompasses studies that uphold the principles of “reliable”, “robust”, “interpretable”, and “explainable” in their methodologies. 

Hence, the following are identified:
\begin{itemize}
    \item Lack of generalizability due to limited data, imbalanced data, biased data.
    \item Lack of robustness to variations in data including diverse imaging conditions, modalities, or populations. 
    \item Lack of explainability in their decision-making process.
    \item Lack of reliability to adopt \gls{cadx} systems in clinical practice.
    \item Lack of appropriate clinical validation across various data and clinical settings. 
\end{itemize}

These gaps highlight the current insufficiencies in CADx systems and underscore the need for future research to not only focus on diagnostic accuracy, but also on enhancing the reliability and transparency of \gls{cadx} systems to facilitate their adoption in clinical practice. Furthermore, the American Association of Physicists in Medicine (AAPM) \cite{aapm2022} has stated that these models are not ready for clinical deployment. A major reason being that these proposed \gls{cadx} systems must undergo proper training and rigorous validation to ensure generalizability and robustness before adoption in clinical practice. Additionally, a few studies have noted that while research in \gls{cadx} systems is increasing, few of these systems are used routinely in clinical settings~\cite{Chan2020,FUTURE-AI_2021}

Rather than focusing solely on improving the predictive performance of glaucoma detection, researchers should prioritize enhancing the generalizability, robustness, explainability, and reliability of the \gls{ai}-based \gls{cadx} systems.  

Consequently, recognizing this gap, we aim to guide the safe design and development of \gls{cadx} systems, with an initial focus on glaucoma diagnosis as part of our future work. This is to ensure that we develop a \gls{cadx} system that is suitable for adoption in clinical practice. Moreover, this framework will include systematic identification of insufficiencies in the model with appropriate mitigation measures to minimize these insufficiencies, ensuring a reliable \gls{cadx} system. Additionally, this framework will be adaptable to various medical imaging and physiological signals applications. Secondly, we will develop a safe and explainable \gls{cadx} system based on this proposed framework, ensuring robust and transparent diagnostic capabilities.

\section{Conclusion}
In conclusion, this paper has provided a comprehensive overview of the \gls{ai} techniques employed in glaucoma diagnosis for \gls{cadx} systems. Through an examination of current research trends, we have identified key gaps and challenges hindering the adoption of \gls{cadx} systems for glaucoma diagnosis. 
This paper serves as a valuable resource in the development and implementation of \gls{cadx} systems. 
As part of our future work, we intend to contribute to the advancement of \gls{cadx} systems especially in regard to safety and reliability and to bring these systems into clinical use.

\bibliographystyle{IEEEtran}
\bibliography{glaucoma}

\end{document}